\documentclass{ws-procs9x6}

\begin{document}

\title{Modeling the galaxy/light-mass  connection with cosmological simulations}

\author{Argyro Tasitsiomi}

\address{Lyman Spitzer Jr. Fellow, Dept. of Astrophysical Sciences, Princeton University,\\
Princeton,  NJ 08544, USA\\
$^*$E-mail: iro@astro.princeton.edu}

\begin{abstract}
I review some results on the galaxy/light-mass connection obtained by  dissipationless simulations in combination
with a simple, non-parametric model to connect halo circular velocity to the luminosity of the galaxy they would host.
I focus on the galaxy-mass correlation and mass-to-light ratios obtained from  galaxy up to cluster scales.
The predictions of this simple scheme are shown to be in very good agreement with SDSS observations. 

\end{abstract}


\bodymatter

\section{Introduction}\label{intro}
During the last decade, large observational galaxy surveys have greatly improved the 
information we have  about the relation of galaxies and/or light with the
underlying mass distribution. However, 
our understanding of these relations is at best incomplete, largely because
they are shaped  by processes of galaxy formation that are too numerous
and complicated to be included in cosmological simulations. As a result, often other either semi-analytic or empirical ways are adopted to investigate these connections.

In this paper I summarize some results on the galaxy/light-dark matter relation predicted by 
a simple scheme used to assign luminosities to dark matter halos formed in cosmological simulations.
In this scheme, simulated dark matter halos of a certain number density  are assigned the luminosity
of observed galaxies with the same number density.  When calculating these number densities, we rank galaxies
with respect to their SDSS $r$-band luminosity --- i.e., we use the observed SDSS $r$-band luminosity function --- 
while we rank the simulated dark matter halos using as tag the maximum of their rotation velocity, $V_{{\rm max}}$.
Thus, the assumption we make is that there is a one-to-one relation between the
luminosity of a galaxy and the gravitational potential of the halo within which the galaxy is formed, with the latter being described
by $V_{\rm max}$. This is a reasonable  assumption since we expect the ability of a halo to make baryons
cool and form a galaxy to increase with the halo gravitational potential depth. 
Furthermore, $V_{max}$ is particularly good as a potential depth 
indicator in the case of subhalos.
Assuming  that the initial --- prior to accretion potential --- is the one relevant for galaxy formation in subhalos, 
$V_{max}$, being a property of the inner parts of halos is not affected as  much by tidal processes as other halo properties (e.g., the mass).
Thus, it is a fair indicator of the initial potential within which the galaxy formed. 
In addition, we assign magnitudes in the remaining four SDSS bands ($u$, $g$, $i$,
and $z$) using the observed relation between local galaxy density and
color. More specifically,  we use as a measure of density the distance to the 10th
nearest neighbor above a certain luminosity.  We measure this quantity for both our
 mock
galaxies and actual observed SDSS galaxies.  For each mock galaxy, we then choose a real SDSS galaxy
which has a similar $r$-band luminosity and nearest neighbor distance,
and assign the colors of this galaxy to the mock galaxy. 

In what follows I present results obtained after assigning luminosities in halos simulated in collisionless simulations 
using the above scheme. More details and references can be found in Refs.~\refcite{tasitsiomi_etal04,mandelbaum_etal05,slosar_etal06,tasitsiomi_etal06}.  
\begin{figure*}
\psfig{file=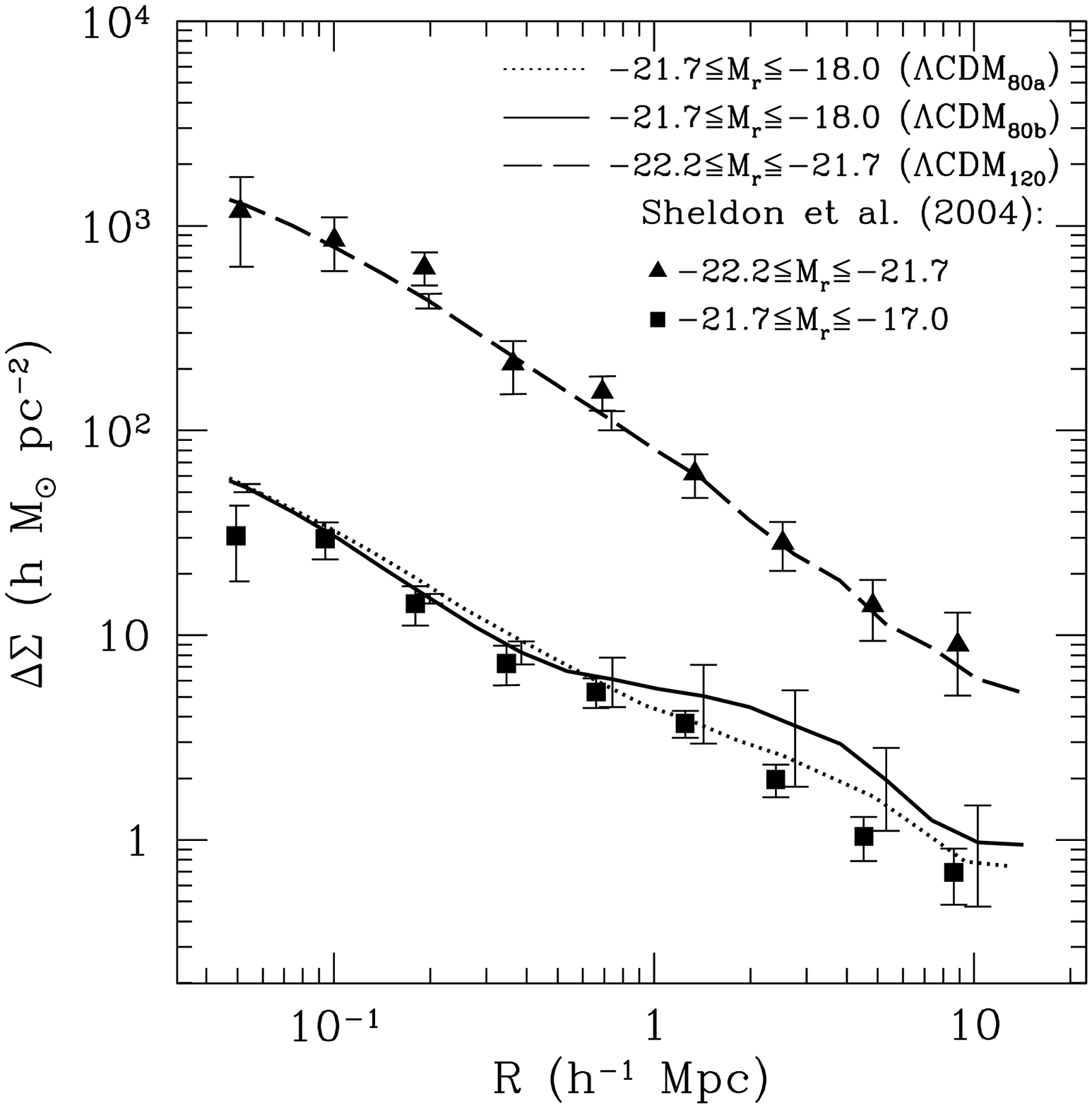,width=2.22in}
\psfig{file=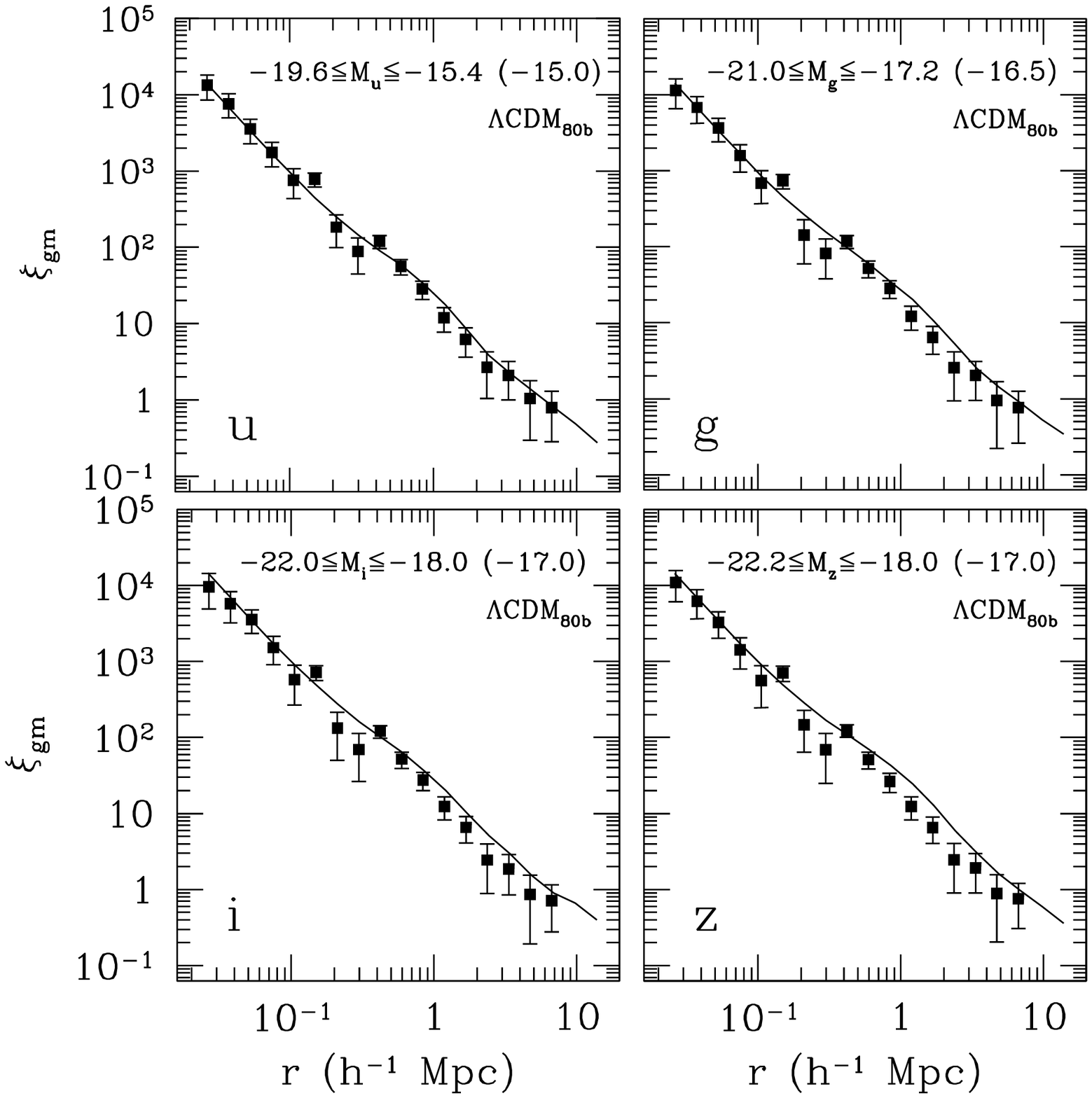,width=2.22in}
\caption{{\it Left panel:} $\Delta \Sigma$ as a function of 
  projected separation $R$ as measured by SDSS\cite{sheldon_etal04} ({\it
    points}) and as calculated from the simulations ({\it lines}).
  Results are shown for two $r$-band luminosity bins: $-22.2\le
  M_{r}\le -21.7$ ({\it triangles/dashed line}) and $-21.7\le M_{r}\le
  -17.0 (-18.0)$ ({\it squares} and {\it solid line}, respectively).
  The effect of $\sigma_8$ can be seen for the fainter sample for
  which we plot the results for the $\sigma_{8}=0.75$ ({\it dotted}
  line) and $\sigma_{8}=0.9$ ({\it solid} line) runs. {\it Right
    panel:} Comparison of the simulation galaxy--mass correlation function 
  ({\it lines}) with the  SDSS observations   
  ({\it points}) in the $u$, $g$, $i$, and $z$ SDSS bands for
  the fainter sample.  The numbers in parentheses denote the actual
  faint magnitude limit of the sample of Ref.~\refcite{sheldon_etal04}.
 Reproduced from Ref.~\refcite{tasitsiomi_etal04}.}
\label{deltas}
\vspace{-0.1cm}
\end{figure*}
\section{Galaxy-mass clustering} 
The most recent observational study of galaxy-galaxy lensing by the
SDSS collaboration \cite{sheldon_etal04} (hereafter S04), has
significantly improved the accuracy of galaxy correlation
measurements.  The observable is $\Delta \Sigma$ defined as
$\Delta \Sigma = \bar{\Sigma} (\le R)-\bar{\Sigma} (R)$,  
where $\bar{\Sigma} (\le R)$ is the mean surface density within the
projected radius $R$, $\bar{\Sigma}(R)$ is the azimuthally averaged
surface density at $R$, and $\Sigma_{\rm crit}$ is the critical density
for lensing which depends on the angular diameter distances of the
lens and the source.  
S04 also deprojected their  $\Delta \Sigma$ to obtain the actual 3D galaxy--mass
correlation function, $\xi_{gm}$.

We calculate $\Delta\Sigma$  and $\xi_{gm}$ in simulations by selecting and weighting objects in accordance with
S04.
 A comparison between our results and the S04 results is shown in Fig.~\ref{deltas}. 
Our faintest bin is
not exactly the same as that of S04, but the faintest galaxies that
are missing from our sample have only a small contribution to the
total signal.
Given these small differences  and the simplicity of our
luminosity assignment scheme the agreement between simulations and observations is quite good.
From the left panel we see that, in agreement with S04, the amplitude of the correlation
function increases with luminosity on intermediate ($0.1-1h^{-1} {\rm Mpc}$)
scales, while it is nearly independent of luminosity on larger scales.
This is indicative of an increase of the effective slope with
luminosity.  
The left panel of the figure also shows the difference in
$\Delta\Sigma(R)$ for two simulations with different values of
$\sigma_8$. In the right panel  we
show the corresponding $\xi_{\rm gm}$ comparisons for the $u$, $g$,
$i$ and $z$ bands  for the faintest S04
sample.  The overall agreement of
our simple luminosity scheme, even in the bluer bandpasses, is
surprisingly good. These results indicate that assigning colors based on luminosity
and environment is sufficient at least for the purposes of galaxy-mass clustering.

\section{Mass-light relation}
\begin{figure}
\begin{center}
\psfig{file=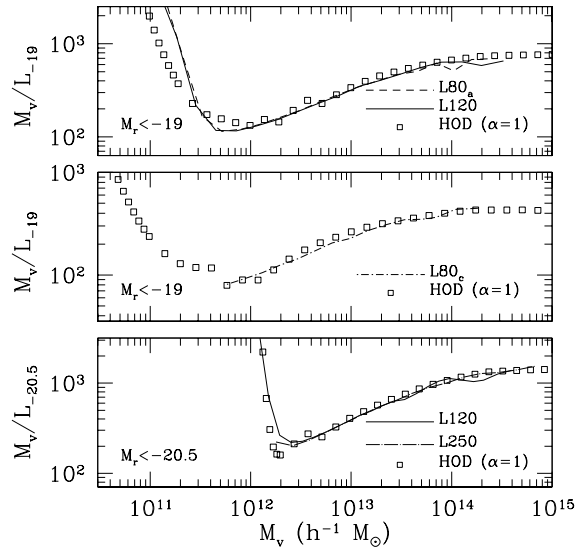,width=3in}
\end{center}
\caption{Mass-to-light ratio against virial mass. For the top two panels
the total luminosity of an object is calculated by summing up the luminosities of all galaxies  with
SDSS $r$-band magnitude $M_{r}<-19$. For the bottom panel this magnitude threshold is equal to $-20.5$. 
 Simulations $L80_{a}, L120$ and $L250$ are for $\Omega_{m}=0.3, \Omega_{\Lambda}=0.3, \sigma_{8}=0.9$ and
$h=0.7$. Simulation $L80_{c}$ was run using the 3 year WMAP best fit parameters. The {\em squares} show the corresponding
HOD predictions of a more recent version of the analysis of Ref.~\refcite{tinker}. In this analysis the authors fixed
the power of the power law giving the mean number of satellite galaxies as 
a function of host mass, $\langle N_{sat} \rangle \propto M^{\alpha}$, to unity ($\alpha =1$). Reproduced from Ref.~\refcite{tasitsiomi_etal06}.}
\label{tinker_test}
\vspace{-0.3cm}
\end{figure}

Our luminosity assignment scheme does not distinguish between isolated halos and subhalos, i.e. halos
within larger (host) halos. Furthermore, there is scatter between mass and $V_{\rm max}$.
Thus, it is not obvious that our scheme can give the observed 
mass-light relation.  In Fig.~\ref{tinker_test} we present a comparison of the mass-to-light ratios as a function of virial mass 
obtained for several of our simulations with results obtained by a more recent version of the M/L analysis of Ref.~\refcite{tinker}.  These authors constrained the
parameters of the Halo Occupation Distribution (HOD) by fitting the space density and the projected SDSS galaxy-galaxy correlation
function. With this HOD they then calculated the mass-to-light ratios. 
The agreement between  our results and those from the HOD analysis 
is astonishing. Furthermore, the mass-to-light ratio varies as expected, namely it has a minimum at some mass scale, and increases towards
larger and smaller masses. It increases more rapidly towards smaller masses, and this increase is expected since below a certain mass one expects
that galaxies cannot form given that halos are not massive enough to be able to cool the baryons. The increase towards high masses is not as steep and
is expected on the basis that  galaxy formation efficiency goes down with mass at large masses possibly because of feedback effects, e.g. from AGNs. 
Moreover, there seems to be a flattening of the mass-to-light ratio above a mass of $\sim 10^{14} h^{-1} M_{\odot}$.   

\section{Going a step further...}
Since our simple scheme seems to be reproducing very well different kinds of observations (also see, e.g., Ref.~\refcite{conroy_etal06} for comparisons with the
galaxy autocorrelation function), we can use it to gain some insight, e.g.,  into interpreting/analyzing the observations. 
\begin{figure*}
\psfig{file=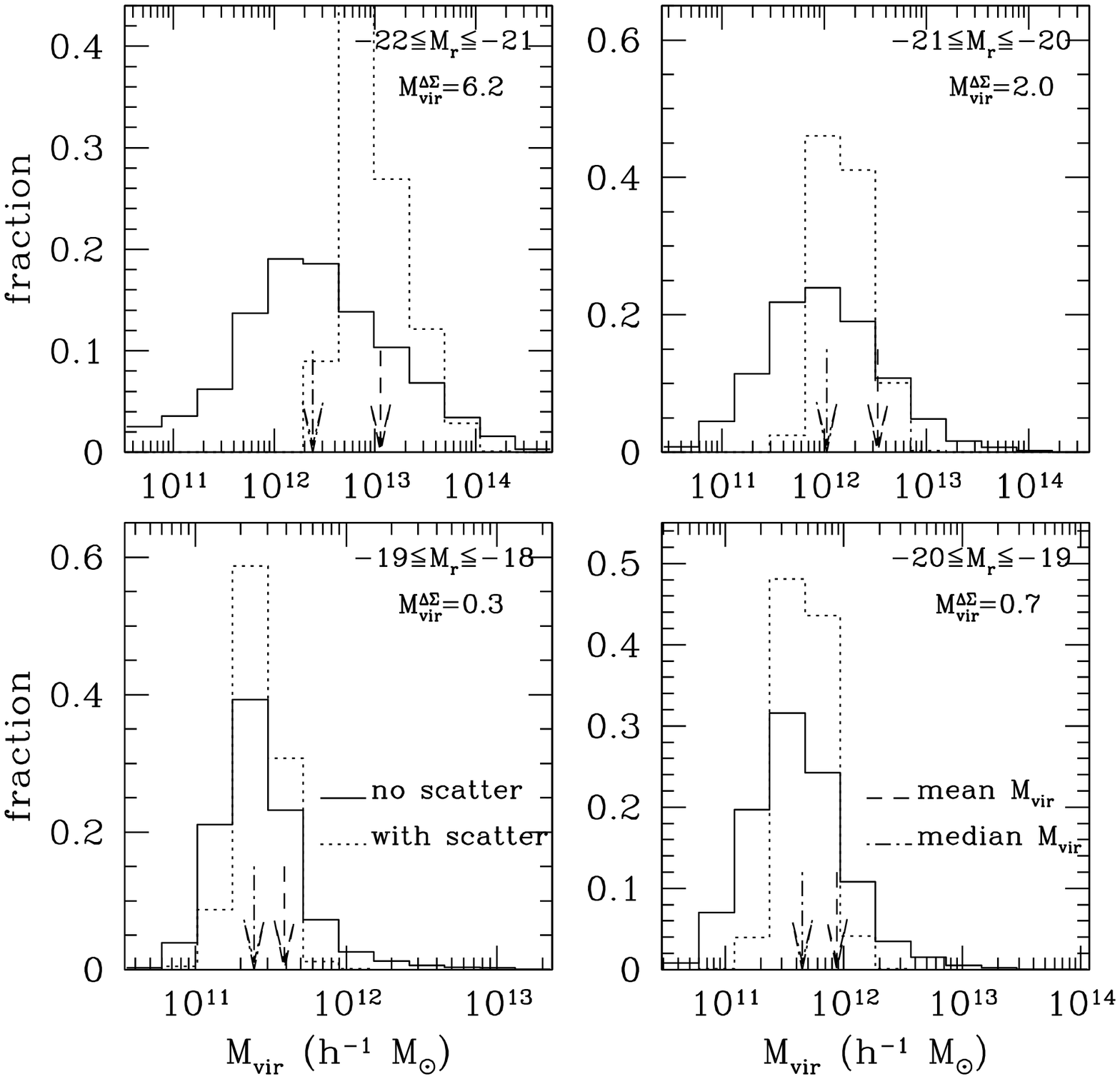,width=2.22in}
\psfig{file=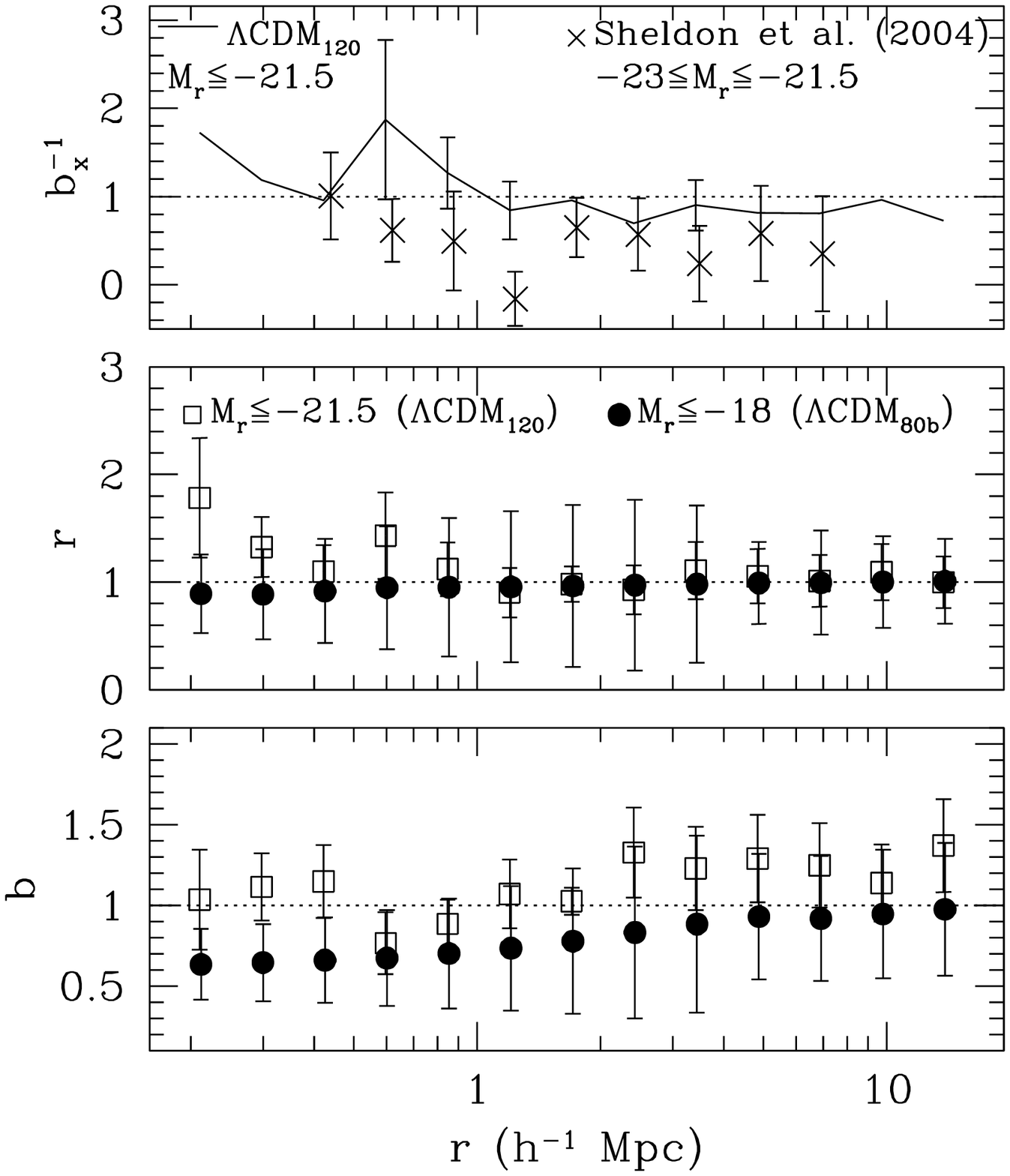,width=2.22in}
\caption{{\it Left panel:} Halo virial mass distributions (defined using an overdensity
  of 180 with respect to the mean density of the universe) for central
  galaxies in four luminosity bins with and without scatter in the
  $V_{max}$-luminosity relation ({\it solid} and {\it dotted}
  histograms, respectively). $M_{\rm vir}^{\Delta \Sigma}$ is the
  virial mass obtained by fitting the $\Delta \Sigma$ of the
  corresponding subsample of galaxies, in units of $10^{12} h^{-1}
  M_{\odot}$.  For the distributions with scatter, the mean and median
  virial masses are indicated with {\it dashed} and {\it dot-dashed}
  arrows, respectively. {\it Right
    panel:}  Bias for the volume-limited samples. {\it Top panel:}  
  Inverse cross-bias, $b_{x}^{-1}$, as measured by SDSS
  \cite{sheldon_etal04} ({\it crosses}) for a sample with $-23\le
  M_{r}\le -21.5$ and as measured in our simulations ({\it solid
    line}) for objects with $M_{r}\le -21.5$.  For clarity, error bars
  in the simulation results are plotted only for scales at which there
  appears to be some discrepancy between simulations and observations.
  {\it Middle panel:} Correlation coefficient as a function of scale
  for two volume-limited simulation samples with $M_{r}\le -18$
  ({\it solid circles}) and $M_{r}\le -21.5$ ({\it open squares}).  {\it Bottom
    panel:} Bias as a function of scale for the same samples as
  in the middle panel. Reproduced from Ref.~\refcite{tasitsiomi_etal04}.}
\label{further}
\vspace{-0.1cm}
\end{figure*}
For example, the left panel of Fig.~\ref{further} addresses the question of how
well weak lensing observations will be able to infer the typical mass of halos in a luminosity
bin. Clearly, for finite luminosity bins as used in observations one has a distribution of halos masses. 
This may bias the estimated 'mean mass', especially if the distribution is asymmetric.  As shown in 
the figure and for our assumed mass distribution, 
we find that the mass inferred via weak lensing $M_{\rm vir}^{\Delta \Sigma}$ lies somewhere between the median
and the mean of the mass distribution.  Thus, the mass derived cannot be interpreted in a
straightforward way as the mass for galaxies of a given luminosity, unless one knows the mass distribution. Therefore, the
mass--luminosity relations should be quoted and interpreted
with caution. 
More discussion on the information we will be able to
extract from weak lensing observations  using the halo model formalism can be found in Ref.~\refcite{mandelbaum_etal05}.

In addition, we can gain insight for quantities not directly observable. The right panel of
Fig.~\ref{further} shows the correlation coefficient $r$ and
bias $b$ for two volume-limited simulation samples of different
luminosity.  The top panel  also shows comparisons of the
cross-bias, $b_x^{-1}$, for the volume-limited sample of S04.  
If $\xi_{mm}, \xi_{gg}$ and $\xi_{gm}$ are the mass--mass, galaxy--galaxy and galaxy--mass
correlation functions, respectively, then the above quantities are defined as:
$b^{2}=\xi_{\rm gg}/\xi_{\rm mm}$, 
$r=\xi_{\rm gm}/({\xi_{\rm mm} \xi_{\rm gg})^{1/2}}$ and
$b_{x}=\xi_{\rm gg}/\xi_{\rm gm}=b/r$.
The
simulation results agree with observations within the error bars.  A
small offset of the simulation results toward smaller values of
$b_{x}$ may be due to objects with $-23.0<M_{r}<-22.5$, which are not
present in our simulations due to the small box size and which could
enhance the clustering signal somewhat.  The correlation coefficient is
approximately unity on scales $\geq 1  h^{-1} {\rm Mpc}$ in our simulations.  This
means that on these scales the cross bias $b_{x}$ which is observable is
a fair measure of the standard bias $b$ which is not directly observable.
\section*{Acknowledgments}
I want to thank A.V. Kravtsov, R.H. Wechsler and J.R. Primack for allowing me to
present here some of the results of our work. Also, I want to especially thank J. Tinker
for making available the predictions of his HOD formalism for comparison with 
the simulation results.

\bibliographystyle{ws-procs9x6}
\bibliography{pro}

\end{document}